\pdfoutput=1
\pdfmapfile{+charter.map}
\pdfmapfile{+mlm.map}

\documentclass[	DIV=calc,%
							paper=a4,%
							fontsize=11pt,
							twocolumn]{scrartcl}	 					

\usepackage{geometry}
\RedeclareSectionCommand[beforeskip=.5\baselineskip]{paragraph}
\usepackage[english]{babel}										
\usepackage[protrusion=true,expansion=true]{microtype}				
\usepackage{enumitem}

\usepackage{amsmath,amsfonts,amsthm}					
\usepackage[pdftex]{graphicx}									
\usepackage[svgnames]{xcolor}									

\usepackage{fancyhdr}												
\usepackage{lastpage}	
\usepackage{tikz}

\usetikzlibrary{calc, shapes.arrows}

\definecolor{myblue}{HTML}{12274d}
\definecolor{mygold}{HTML}{a85c04}
\definecolor{mygrey}{HTML}{A3A9AA}
\definecolor{Indigo}{HTML}{500472} 
\definecolor{Bole}{HTML}{784F41} 
\definecolor{Tiffany}{HTML}{79cbb8}
\definecolor{Isabelline}{HTML}{F1E8E4}

\definecolor{Chocolate}{HTML}{420C14} 
\definecolor{PolyGreen}{HTML}{214e34}
\definecolor{Tiger}{HTML}{A85C04} 
\definecolor{Ivory}{HTML}{EEF3E8}
\definecolor{Puce}{HTML}{C08497}
\definecolor{YInMn}{HTML}{26547C}
\definecolor{Gamboge}{HTML}{F06543}
\colorlet{mygray}{mygrey!75!black}
\usepackage{afterpage}

\colorlet{titlecolor}{myblue}
\colorlet{linecolor}{mygold}
\colorlet{affcolor}{mygray}
\colorlet{titlecolor}{Black}
\colorlet{linecolor}{Grey}
\colorlet{affcolor}{Grey}
\pagecolor{Ivory!30}

\usepackage{charter} 
\def\fonttitle{LinuxBiolinumT-OsF}

\def\fontauthors{LinuxBiolinumT-OsF}

\newcommand{\formatauthor}[1]{
\fontfamily{\fontauthors}\selectfont
\color{titlecolor}#1}
\newcommand{\formataffil}[1]{&\footnotesize \fontfamily{\fontauthors}\selectfont\color{affcolor}#1\\}
\usepackage{titling}															
\usepackage{array}

\newcommand{\HorRule}{\color{linecolor}
									  	\rule{\linewidth}{1pt}%
										}
\pretitle{\vspace{-52pt}%
\begin{minipage}[b]{0.163\linewidth}
\includegraphics[width=\linewidth]{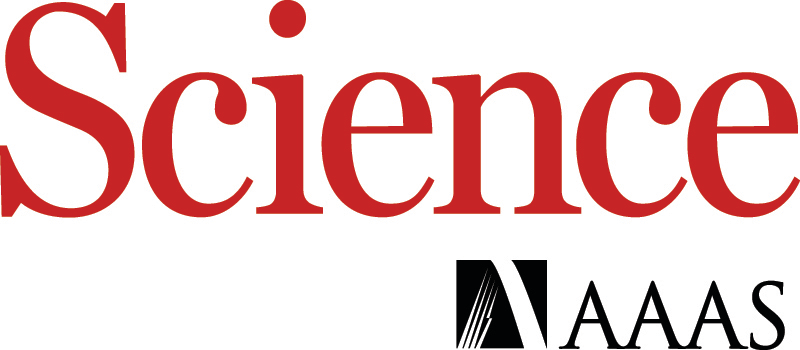}
\end{minipage}\hfill\begin{minipage}[b]{0.817\linewidth}
{\small This is the author's version of the work. It is posted here by permission of the AAAS for personal use, not for redistribution. The definitive version was published in Science on 20 May 2024, DOI: \href{https://www.science.org/doi/10.1126/science.adn0117}{10.1126/science.adn0117}.}\end{minipage}
\begin{center}\vspace{1em} \HorRule\\[1em] 
				\fontsize{26}{30}\fontfamily{\fonttitle}\selectfont
    \bfseries
    \color{titlecolor} 
				}
\title{Managing extreme AI risks \\
amid rapid progress}					
\posttitle{\par\end{center}\vskip 0.5em}

\preauthor{\hspace{1.3cm}\begin{tabular}[b]{m{3.5cm}m{0.65\linewidth}}
}

\author{
\formatauthor{Yoshua Bengio}
\formataffil{Mila - Quebec AI Institute, Université de Montréal}
\formatauthor{Geoffrey Hinton}
\formataffil{University of Toronto, Vector Institute}
\formatauthor{Andrew Yao}
\formataffil{Tsinghua University}
\formatauthor{Dawn Song}
\formataffil{UC Berkeley}
\formatauthor{Pieter Abbeel}
\formataffil{UC Berkeley}
\formatauthor{Trevor Darrell}
\formataffil{UC Berkeley}
\formatauthor{Yuval Noah Harari}
\formataffil{The Hebrew University of Jerusalem}
\formatauthor{Ya-Qin Zhang}
\formataffil{Tsinghua University}
\formatauthor{Lan Xue}
\formataffil{Institute for AI International Governance, Tsinghua University}
\formatauthor{Shai Shalev-Shwartz}
\formataffil{The Hebrew University of Jerusalem}
\formatauthor{Gillian Hadfield}
\formataffil{University of Toronto, Schwartz Reisman Inst. for Technology and Society, Vector Inst.}
\formatauthor{Jeff Clune}
\formataffil{University of British Columbia, Vector Institute}
\formatauthor{Tegan Maharaj}
\formataffil{University of Toronto, Schwartz Reisman Inst. for Technology and Society, Vector Inst.}
\formatauthor{Frank Hutter}
\formataffil{ELLIS Institute Tübingen, University of Freiburg}
\formatauthor{Atılım Güneş Baydin}
\formataffil{University of Oxford}
\formatauthor{Sheila McIlraith}
\formataffil{University of Toronto, Schwartz Reisman Inst. for Technology and Society, Vector Inst.}
\formatauthor{Qiqi Gao}
\formataffil{East China University of Political Science and Law}
\formatauthor{Ashwin Acharya}
\formataffil{RAND Corporation}
\formatauthor{David Krueger}
\formataffil{University of Cambridge}
\formatauthor{Anca Dragan}
\formataffil{UC Berkeley}
\formatauthor{Philip Torr}
\formataffil{University of Oxford}
\formatauthor{Stuart Russell}
\formataffil{UC Berkeley}
\formatauthor{Daniel Kahneman}
\formataffil{School of Public and International Affairs, Princeton University}
\formatauthor{Jan Brauner*}
\formataffil{University of Oxford, RAND Corporation}
\formatauthor{Sören Mindermann*}
\formataffil{University of Oxford, Mila - Quebec AI Institute, Université de Montréal}
}											
\postauthor{ 								
					\end{tabular}\par\vspace{2mm}\HorRule\vspace*{-10mm}}     
\date{}																				

\usepackage[backend=biber, bibstyle=ieee, citestyle=numeric-comp,
  sorting=none, labeldateparts,
  maxbibnames=99, url=false, isbn=false, maxcitenames=2, mincitenames=1]{biblatex}  
\PassOptionsToPackage{hyphens}{url}
\AtEveryBibitem{\clearlist{language}}
\renewbibmacro*{date}{%
  \iffieldundef{year}
    {\bibstring{nodate}}
    {\printdate}}

\usepackage{hyperref}
\hypersetup{
colorlinks   = true, 
linkcolor=black,
citecolor=titlecolor,
urlcolor=mygrey!75!black,
}
\usepackage{xurl}

\begin{document}
\makeatletter
\twocolumn[
   \begin{@twocolumnfalse} 
     \maketitle 
     \begin{abstract}
     \vspace{-4mm}
     \subsection*{\hspace{0.445\linewidth}{Abstract}}
     
     \noindent Artificial Intelligence (AI) is progressing rapidly, and companies are shifting their focus to developing generalist AI systems that can autonomously act and pursue goals. Increases in capabilities and autonomy may soon massively amplify AI’s impact, with risks that include large-scale social harms, malicious uses, and an irreversible loss of human control over autonomous AI systems. Although researchers have warned of extreme risks from AI \supercite{noauthor_2023-np}, there is a lack of consensus about how exactly such risk arise, and how to manage them.  Society’s response, despite promising first steps, is incommensurate with the possibility of rapid, transformative progress that is expected by many experts. AI safety research is lagging. Present governance initiatives lack the mechanisms and institutions to prevent misuse and recklessness, and barely address autonomous systems. In this short consensus paper, we describe extreme risks from upcoming, advanced AI systems. Drawing on lessons learned from other safety-critical technologies, we then outline a comprehensive plan combining technical research and development (R\&D) with proactive, adaptive governance mechanisms for a more commensurate preparation.
     \end{abstract}
     \vspace{-13mm}
    \end{@twocolumnfalse}
]
\makeatother
\newpage
\thispagestyle{fancy} 			
\vfill\null\newpage
\subsection*{Rapid progress}

\noindent
Current deep learning systems still lack important capabilities and we do not know how long it will take to develop them. However, companies are engaged in a race to create generalist AI systems that match or exceed human abilities in most cognitive work \supercite{DeepMind_undated-dk,OpenAI_undated-sz}. They are rapidly deploying more resources and developing new techniques to increase AI capabilities, with investment in training state-of-the-art models tripling annually \supercite{Cottier2023-ko}.

There is much room for further advances, as tech companies have the cash reserves needed to scale the latest training runs by multiples of 100 to 1000 \supercite{Alphabet2022-ro}. Hardware and algorithms will also improve: AI computing chips have been getting 1.4 times more cost-effective, and AI training algorithms 2.5 times more efficient, each year \supercite{Hobbhahn2023-wl,Erdil2022-bc}. Progress in AI also enables faster AI progress \supercite{noauthor_undated-tf}: AI assistants are increasingly used to automate programming \supercite{Tabachnyk_undated-kq}, data collection \supercite{Bai2022-sv,OpenAI2023-bl}, and chip design \supercite{Mirhoseini2021-zs}.

There is no fundamental reason for AI progress to slow or halt at human-level abilities. Indeed, AI has already surpassed human abilities in narrow domains like playing strategy games and predicting how proteins fold \supercite{Jumper2021-jy,Brown2019-pw,Campbell2002-dt}. Compared to humans, AI systems can act faster, absorb more knowledge, and communicate at higher bandwidth. Additionally, they can be scaled to use immense computational resources and can be replicated by the millions.

We don’t know for certain how the future of AI will unfold. However, we must take seriously the possibility that highly powerful generalist AI systems---outperforming human abilities across many critical domains---will be developed within the current decade or the next. What happens then?

More capable AI systems have larger impacts. Especially as AI matches and surpasses human workers in capabilities and cost-effectiveness, we expect a massive increase in AI deployment, opportunities, and risks. If managed carefully and distributed fairly, AI could help humanity cure diseases, elevate living standards, and protect ecosystems. The opportunities are immense.

But alongside advanced AI capabilities come large-scale risks that we are not on track to handle well. Humanity is pouring vast resources into making AI systems more powerful but far less into their safety and mitigating their harms. Only an estimated 1-3\% of AI publications are on safety \supercite{Toner2022-if,Emerging_Technology_Observatory_undated-ix}. For AI to be a boon, we must reorient; pushing AI capabilities alone is not enough.

We are already behind schedule for this reorientation. The scale of the risks means that we need to be proactive, as the costs of being unprepared far outweigh those of premature preparation. We must anticipate the amplification of ongoing harms, as well as novel risks, and prepare for the largest risks well before they materialize. Climate change has taken decades to be acknowledged and confronted; for AI, decades could be too long.

\subsection*{Societal-scale risks}\label{high-stakes-risks}

If not carefully designed and deployed, increasingly advanced AI systems threaten to amplify social injustice, erode social stability, and weaken our shared understanding of reality that is foundational to society. They could also enable large-scale criminal or terrorist activities. Especially in the hands of a few powerful actors, AI could cement or exacerbate global inequities, or facilitate automated warfare, customized mass manipulation, and pervasive surveillance \supercite{Weidinger2022-he,Chan2023-na,Eubanks2018-hr,Hendrycks2023-zc,Bommasani2021-bb,Solaiman2023-uo}. 

Many of these risks could soon be amplified, and new risks created, as companies are working to develop autonomous AI: systems that can pursue goals and act in the world. While current AI systems have limited autonomy, work is underway to change this \supercite{Wang2023-tq}. For example, the non-autonomous GPT-4 model was quickly adapted to browse the web, design and execute chemistry experiments, and utilize software tools, including other AI models \supercite{noauthor_undated-wh,Bran2023-fu,Mialon2023-ri,Shen2023-xy}.

If we build highly advanced autonomous AI, we risk creating systems that pursue undesirable goals. Malicious actors could deliberately embed undesirable goals. Without R\&D breakthroughs (see below), even well-meaning developers may inadvertently create AI systems pursuing unintended goals: The reward signal used to train AI systems usually fails to fully capture the intended objectives, leading to AI systems that pursue the literal specification rather than the intended outcome \supercite{hadfield2019incomplete}. Additionally, the training data never captures all relevant situations, leading to AI systems that pursue undesirable goals in novel situations encountered after training.

Once autonomous AI systems pursue undesirable goals, we may be unable to keep them in check. Control of software is an old and unsolved problem: computer worms have long been able to proliferate and avoid detection \supercite{Denning1989-ds}. However, AI is making progress in critical domains such as hacking, social manipulation, and strategic planning \supercite{Wang2023-tq,Park2023-db}, and may soon pose unprecedented control challenges.

To advance undesirable goals, future autonomous AI systems could use undesirable strategies---learned from humans or developed independently---as a means to an end \supercite{Turner2019-pz,Perez2022-is,Pan2023-zb,Hadfield-Menell2017-lq}. AI systems could gain human trust, acquire financial resources, influence key decision-makers, and form coalitions with human actors and other AI systems. To avoid human intervention \supercite{Hadfield-Menell2017-lq}, they might copy their algorithms across global server networks \supercite{Kinniment2023-oz}, as computer worms do. AI assistants are already co-writing a substantial share of computer code worldwide \supercite{Dohmke_undated-cc}; future AI systems could insert and then exploit security vulnerabilities to control the computer systems behind our communication, media, banking, supply-chains, militaries, and governments. In open conflict, AI systems could autonomously deploy a variety of weapons, including biological ones. AI systems having access to such technology would merely continue existing trends to automate military activity and biological research. If AI systems pursued such strategies with sufficient skill, it would be difficult for humans to intervene.

Finally, AI systems will not need to plot for influence if it is freely handed over. As autonomous AI systems increasingly become faster and more cost-effective than human workers, a dilemma emerges. Companies, governments, and militaries might be forced to deploy AI systems widely and cut back on expensive human verification of AI decisions, or risk being outcompeted \supercite{Chan2023-na,Critch2023-iq}. As a result, autonomous AI systems could increasingly assume critical societal roles.

Without sufficient caution, we may irreversibly lose control of autonomous AI systems, rendering human intervention ineffective. Large-scale cybercrime, social manipulation, and other harms could escalate rapidly. This unchecked AI advancement could culminate in a large-scale loss of life and the biosphere, and the marginalization or extinction of humanity.

Harms such as misinformation and discrimination from algorithms are already evident today; other harms show signs of emerging. It is vital to both address ongoing harms and anticipate emerging risks. This is not a question of either/or. Present and emerging risks often share similar mechanisms, patterns, and solutions \supercite{Brauner2023-jr}; investing in governance frameworks and AI safety will bear fruit on multiple fronts \supercite{Center-for-AI-Safety2023-jj}.

\subsection*{Reorient Technical R\&D}\label{a-path-forward}

There are many open technical challenges in ensuring the safety and ethical use of generalist, autonomous AI systems. Unlike advancing AI capabilities, these challenges cannot be addressed by simply using more computing power to train bigger models. They are unlikely to resolve automatically as AI systems get more capable \supercite{McKenzie2023-ny,Pan2022-dt,Casper2023-ll,Hendrycks2021-vy,Perez2022-is,Wei2023-lp}, and require dedicated research and engineering efforts. In some cases, leaps of progress may be needed; we thus do not know if technical work can fundamentally solve these challenges in time. However, there has been comparatively little work on many of these challenges. More R\&D may thus make progress and reduce risks.

A first set of R\&D areas needs breakthroughs to enable reliably safe AI. Without this progress, developers must either risk creating unsafe systems or falling behind competitors who are willing to take more risks. If ensuring safety remains too difficult, extreme governance measures would be needed to prevent corner-cutting driven by competition and overconfidence.  These R\&D challenges include:

\paragraph{Oversight and honesty:} More capable AI systems can better exploit weaknesses in technical oversight and testing \supercite{Pan2022-dt,Zhuang2020-jn,Gao2023-bd}---for example, by producing false but compelling output \supercite{Sharma2023-yb,Amodei_undated-zo,Casper2023-ll}.

\paragraph{Robustness:} AI systems behave unpredictably in new situations. While some aspects of robustness improve with model scale \supercite{Hendrycks2020-zt}, other aspects do not or even get worse \supercite{noauthor_undated-xr,Shah2022-ej,Hendrycks2021-vy,Wang2022-iv}.

\paragraph{Interpretability and transparency:} AI decision-making is opaque, with larger, more capable, models being more complex to interpret. So far, we can only test large models via trial and error. We need to learn to understand their inner workings \supercite{Rauker2023-jm}.

\paragraph{Inclusive AI development:} AI advancement will need methods to mitigate biases and integrate the values of the many populations it will affect \supercite{Eubanks2018-hr,Sen1986-ym}.

\paragraph{Addressing emerging challenges:} Future AI systems may exhibit failure modes we have so far seen only in theory or lab experiments, such as AI systems taking control over the training reward-provision channels or exploiting weaknesses in our safety objectives and shutdown mechanisms to advance a particular goal \supercite{Hadfield-Menell2017-lq,Ngo2024-dz,Hubinger2024-mw,Cohen2024-am}.

A second set of R\&D challenges need progress to enable effective, risk-adjusted governance, or reduce harms when safety and governance fail:

\paragraph{Evaluation for dangerous capabilities:} As AI developers scale their systems, unforeseen capabilities appear spontaneously, without explicit programming \supercite{Wei2022-pa}. They are often only discovered after deployment \supercite{Wei2022-vg,Zhou2024-uu,Davidson2023-tp}. We need rigorous methods to elicit and assess AI capabilities, and to predict them before training. This includes both generic capabilities to achieve ambitious goals in the world (e.g., long-term planning and execution), as well as specific dangerous capabilities based on threat models (e.g. social manipulation or hacking). Current evaluations of frontier AI models for dangerous capabilities \supercite{Shevlane2023-po}---key to various AI policy frameworks---are limited to spot-checks and attempted demonstrations in specific settings \supercite{Kinniment2023-oz,Mouton2024-ap,Scheurer2023-yj}. These evaluations can sometimes demonstrate dangerous capabilities but cannot reliably rule them out: AI systems that lacked certain capabilities in the tests may well demonstrate them in slightly different settings or with post-training enhancements. Decisions that depend on AI systems not crossing any red lines thus need large safety margins. Improved evaluation tools decrease the chance of missing dangerous capabilities, allowing for smaller margins.

\paragraph{Evaluating AI alignment:} If AI progress continues, AI systems will eventually possess highly dangerous capabilities. Before training and deploying such systems, we need methods to assess their propensity to use these capabilities. Purely behavioral evaluations may fail for advanced AI systems: like humans, they might behave differently under evaluation, faking alignment \supercite{Hubinger2024-mw,Ngo2024-dz,Cohen2024-am}.

\paragraph{Risk assessment:} We must learn to assess not just dangerous capabilities, but risk in a societal context, with complex interactions and vulnerabilities. Rigorous risk assessment for frontier AI systems remains an open challenge due to their broad capabilities and pervasive deployment across diverse application areas \supercite{Koessler2023-uq}. 

\paragraph{Resilience:} Inevitably, some will misuse or act recklessly with AI. We need tools to detect and defend against AI-enabled threats such as large-scale influence operations, biological risks, and cyber-attacks. However, as AI systems become more capable, they will eventually be able to circumvent human-made defenses. To enable more powerful AI-based defenses, we first need to learn how to make AI systems safe and aligned.

Given the stakes, we call on major tech companies and public funders to allocate at least one-third of their AI R\&D budget---comparable to their funding for AI capabilities---towards addressing the above R\&D challenges and ensuring AI safety and ethical use \supercite{Hendrycks2021-vy}. Beyond traditional research grants, government support could include prizes, advance market commitments \supercite{Ho2021-ac}, and other incentives. Addressing these challenges, with an eye toward powerful future systems, must become central to our field.

\subsection*{Governance measures}

We urgently need national institutions and international governance to enforce standards preventing recklessness and misuse. Many areas of technology, from pharmaceuticals to financial systems and nuclear energy, show that society requires and effectively uses government oversight to reduce risks. However, governance frameworks for AI are far less developed, lagging behind rapid technological progress. We can take inspiration from the governance of other safety-critical technologies, while keeping the uniqueness of advanced AI in mind: that it far outstrips other technologies in its potential to act and develop ideas autonomously, progress explosively, behave adversarially, and cause irreversible damage.

Governments worldwide have taken positive steps on frontier AI, with key players including China, the US, the EU, and the UK engaging in discussions \supercite{AI_Safety_Summit2023-oe,Oecd2023-ic} and introducing initial guidelines or regulations \supercite{The_White_House_US2023-dk,Cyberspace_Administration_of_China2023-mz,European_Union2024-pk,Department_of_State_for_Science_Innovation_and_Technology_UK2023-oq}. Despite their limitations---often voluntary adherence, limited geographic scope, and exclusion of high-risk areas like military and R\&D-stage systems---they are important initial steps towards, amongst others, developer accountability, third-party audits, and industry standards.

Yet, these governance plans fall critically short in view of the rapid progress in AI capabilities. We need governance measures that prepare us for sudden AI breakthroughs, while being politically feasible despite disagreement and uncertainty about AI timelines. The key is policies that automatically trigger when AI hits certain capability milestones. If AI advances rapidly, strict requirements automatically take effect, but if progress slows, the requirements relax accordingly. Rapid, unpredictable progress also means that risk reduction efforts must be proactive—identifying risks from next-generation systems and requiring developers to address them before taking high-risk actions. We need fast-acting, tech-savvy institutions for AI oversight, mandatory and much more rigorous risk assessments with enforceable consequences (including assessments that put the burden of proof on AI developers), and mitigation standards commensurate to powerful autonomous AI. 

Without these, companies, militaries, and governments may seek a competitive edge by pushing AI capabilities to new heights while cutting corners on safety, or by delegating key societal roles to autonomous AI systems with insufficient human oversight; reaping the rewards of AI development while leaving society to deal with the consequences.

\paragraph{Institutions to govern the rapidly moving frontier of AI.} To keep up with rapid progress and avoid quickly outdated, inflexible laws \supercite{Xue_undated-na,Maas_undated-oi,Xue2019-sm} national institutions need strong technical expertise and the authority to act swiftly. To facilitate technically demanding risk assessments and mitigations, they will require far greater funding and talent than they are due to receive under almost any current policy plan. To address international race dynamics, they need the affordance to facilitate international agreements and partnerships \supercite{Ho2023-bb,Trager2023-rm}. Institutions should protect low-risk use and low-risk academic research, by avoiding undue bureaucratic hurdles for small, predictable AI models. The most pressing scrutiny should be on AI systems at the frontier: the few most powerful systems – trained on billion-dollar supercomputers – which will have the most hazardous and unpredictable capabilities \supercite{Anderljung2023-mu,Ganguli2022-no}.

\paragraph{Government insight.} To identify risks, governments urgently need comprehensive insight into AI development. Regulators should mandate whistleblower protections, incident reporting, registration of key information on frontier AI systems and their data sets throughout their life cycle, and monitoring of model development and supercomputer usage \supercite{Kolt2024-im}. Recent policy developments should not stop at requiring that companies report the results of voluntary or underspecified model evaluations shortly before deployment \supercite{The_White_House_US2023-dk,European_Union2024-pk}. Regulators can and should require that frontier AI developers grant external auditors on-site, comprehensive (“white-box”), and fine-tuning access from the start of model development \supercite{Casper2024-tn}. This is needed to identify dangerous model capabilities such as autonomous self-replication, large-scale persuasion, breaking into computer systems, developing (autonomous) weapons, or making pandemic pathogens widely accessible \supercite{Phuong2024-hs,Shevlane2023-po,Mokander2023-pa,Mouton2024-ap,Scheurer2023-yj,Kinniment2023-oz}.

\paragraph{Safety cases.} Despite evaluations, we cannot consider coming powerful frontier AI systems “safe unless proven unsafe”. With current testing methodologies, issues can easily be missed. Additionally, it is unclear if governments can quickly build the immense expertise needed for reliable technical evaluations of AI capabilities and societal-scale risks. Given this, developers of frontier AI should carry the burden of proof to demonstrate that their plans keep risks within acceptable limits. Doing so, they would follow best practices for risk management from industries such as aviation \supercite{European_Organisation_for_the_Safety_of_Air_Navigation2010-bf}, medical devices \supercite{Food_and_Drug_Administration2014-ze}, and defense software \supercite{noauthor_2023-pa}, where companies make safety cases \supercite{Clymer2024-jo,Kelly2004-bq,Mcdermid2020-qk,Isoiec2023-nd,Raz2005-ws}: structured arguments with falsifiable claims supported by evidence, which identify potential hazards, describe mitigations, show that systems will not cross certain red lines, and model possible outcomes to assess risk. Safety cases could leverage developers’ in-depth experience with their own systems. Safety cases are politically viable even when people disagree on how advanced AI will become, since it is easier to demonstrate a system is safe when its capabilities are limited. Governments are not passive recipients of safety cases: they set risk thresholds, codify best practices, employ experts and third-party auditors to assess safety cases and conduct independent model evaluations, and hold developers liable if their safety claims are later falsified.

\paragraph{Mitigation.} To keep AI risks within acceptable limits, we need governance mechanisms matched to the magnitude of the risks \supercite{Anderljung2023-mu,AI_Now_Institute_undated-lx,Schuett2023-ad,Hadfield2023-eb}. Regulators should clarify legal responsibilities arising from existing liability frameworks and hold frontier AI developers and owners legally accountable for harms from their models that can be reasonably foreseen and prevented---including harms that foreseeably arise from deploying powerful AI systems whose behavior they cannot predict. Liability, together with consequential evaluations and safety cases, can prevent harm and create much-needed incentives to invest in safety.

Commensurate mitigations are needed for exceptionally capable future AI systems, like autonomous systems that could circumvent human control. Governments must be prepared to license their development, restrict their autonomy in key societal roles, halt their development and deployment in response to worrying capabilities, mandate access controls, and require information security measures robust to state-level hackers, until adequate protections are ready. Governments should build these capacities now.

To bridge the time until regulations are complete, major AI companies should promptly lay out “if-then” commitments: specific safety measures they will take if specific red-line capabilities \supercite{Shevlane2023-po} are found in their AI systems. These commitments should be detailed and independently scrutinized. Regulators should encourage a race-to-the-top among companies by using the best-in-class commitments, together with other inputs, to inform standards that apply to all players.

To steer AI toward positive outcomes and away from catastrophe, we need to reorient. There is a responsible path, if we have the wisdom to take it.

\subsection*{Acknowledgments}

Yoshua Bengio, Jeff Clune, Gillian Hadfield, Sheila McIlraith hold the position of CIFAR AI Chair. Jeff Clune is a Senior Research Advisor to Google DeepMind. Ashwin Acharya reports acting as an advisor to the Civic AI Security Program. Ashwin Acharya was affiliated with the Institute for AI Policy and Strategy at the time of the first submission. Anca Dragan now hold an appointment at Google DeepMind, but joined the company after the manuscript was written. Dawn Song is the president of Oasis Labs. Trevor Darrell is a cofounder of Prompt AI. Pieter Abbeel is a cofounder at covariant.ai and Investment Partner at AIX Ventures. Shai Shalev-Shwartz is the CTO at Mobileye. David Krueger served as a Research Director for the UK Foundation Model Task Force in 2023, and joined the board of the non-profit Center for AI Policy in 2024. Gillian Hadfield reports the following activities: 2018-2023: Senior Policy Advisor, OpenAI; 2023-present: Member, RAND Technology Advisory 
Group; 2022-present: Member, Partnership on AI, Safety Critical AI Steering Committee. In gratitude and remembrance of Daniel Kahneman, our co-author, whose remarkable contributions to this paper and to humanity's cumulative knowledge and wisdom will never be forgotten.

\subsection*{References and notes}

\printbibliography[heading=none]

\end{document}